\documentclass[twocolumn,aps,prd,showpacs,superscriptaddress]{revtex4-1}

\usepackage{graphicx}
\usepackage{amssymb}
\usepackage{amsmath}
\usepackage{color}

\begin{document}

\title{Multi-photon Effects in Single Nonlinear Compton Scattering and Single Nonlinear Breit-Wheeler Process in Ultra Intense Fields}

\author{Bo Zhang}%
\email{zhangbolfrc@caep.cn}
\affiliation{Department of High Energy Density Physics, Research Center of Laser Fusion, 621900, Mianyang, Sichuan, People's Republic of China}%
\affiliation{Laboratory of Science and Technology on Plasma Physics, Research Center of Laser Fusion, 621900, Mianyang, Sichuan, People's Republic of China}%

\author{Zhi-meng Zhang}
\affiliation{Department of High Energy Density Physics, Research Center of Laser Fusion, 621900, Mianyang, Sichuan, People's Republic of China}%
\affiliation{Laboratory of Science and Technology on Plasma Physics, Research Center of Laser Fusion, 621900, Mianyang, Sichuan, People's Republic of China}%

\author{Zhi-gang Deng}
\affiliation{Department of High Energy Density Physics, Research Center of Laser Fusion, 621900, Mianyang, Sichuan, People's Republic of China}%

\author{Wei Hong,}
\affiliation{Department of High Energy Density Physics, Research Center of Laser Fusion, 621900, Mianyang, Sichuan, People's Republic of China}%
\affiliation{Laboratory of Science and Technology on Plasma Physics, Research Center of Laser Fusion, 621900, Mianyang, Sichuan, People's Republic of China}%

\author{Jian Teng}
\affiliation{Department of High Energy Density Physics, Research Center of Laser Fusion, 621900, Mianyang, Sichuan, People's Republic of China}%
\affiliation{Laboratory of Science and Technology on Plasma Physics, Research Center of Laser Fusion, 621900, Mianyang, Sichuan, People's Republic of China}%

\author{Shu-kai He}
\affiliation{Department of High Energy Density Physics, Research Center of Laser Fusion, 621900, Mianyang, Sichuan, People's Republic of China}%

\author{Wei-min Zhou}
\affiliation{Department of High Energy Density Physics, Research Center of Laser Fusion, 621900, Mianyang, Sichuan, People's Republic of China}%
\affiliation{Laboratory of Science and Technology on Plasma Physics, Research Center of Laser Fusion, 621900, Mianyang, Sichuan, People's Republic of China}%

\author{and Yu-qiu Gu}
\email{yqgu@caep.cn}
\affiliation{Department of High Energy Density Physics, Research Center of Laser Fusion, 621900, Mianyang, Sichuan, People's Republic of China}%
\affiliation{Laboratory of Science and Technology on Plasma Physics, Research Center of Laser Fusion, 621900, Mianyang, Sichuan, People's Republic of China}%

\date{\today}

\begin{abstract}
        Nonlinear Compton scattering (NCS) and nonlinear Breit-Wheeler (NBW) process are strongly multi-photon and highly nonlinear processes. In ultra intense lasers (normalized field amplitude $a_0 \gg 1$), radiation formation length is much shorter than a period and single NCS/NBW cannot be described as scatterings of electrons dressing plane waves with $\gamma$ photons for what they feel is a local constant crossed field. However, present theories in constant crossed fields are hard to give some important quantum features due to divergence problems, such as number of laser photons involved, instantaneous angular distribution and detailed spectrum. As an alternative, present understanding of single NCS/NBW in ultra intense lasers includes several classical and semi-quantum ideas such as forward emission, recoil reaction and spectrum cutoff. We investigated multi-photon effects on NCS/NBW in ultra intense lasers by extracting the number of laser photons involved in a single process in ultra intense lasers from formulae of existing theories. New features of single NCS in ultra intense lasers including fixed emission angle to instantaneous electron momentum, instantaneous deflection of electron, and disappearance of spectrum cutoff are deduced. Similar features of single NBW in ultra intense lasers including non-vanishing emission angles to instantaneous $\gamma$ photon momentum, disappearance of spectrum cutoff and appearance of spectrum lower limit are also obtained. Simulations show that corresponding signals of multi-photon effects are significant on $10$PW scale and stronger lasers.
\end{abstract}


\maketitle

\section{Introduction}\label{sec:intro}
    Strong field quantum electrodynamics (SFQED) attracted a surge of interest recently. With the presence of a strong external field, many QED processes that forbidden or highly suppressed in vacuum would become important, such as Schwinger pair production \cite{Sauter1931,Heisenberg1936,Schwinger1951}, vacuum birefringence \cite{VacuumBirefringence}, Unruh radiation \cite{Unruh1976,Schuetzhold2006}, photon splitting \cite{PhotonSplitting}, light bending \cite{Kim2012}, vacuum radiation \cite{VacRad} and enhanced neutrino-photon coupling \cite{Shaisultanov2000}. For recent reviews, see \cite{ReviewDiPiazza2012,ReviewEhlotzky2009,ReviewMarklund2006}.

    Laser is presently the most intense electromagnetic field in laboratory and some SFQED experiments are expected to be carried out on laser facilities in foreseeable future. So far, the highest laser intensity reported is $2\times 10^{22}$W/cm$^2$ \cite{Yanovsky2008}, which corresponds to $a_0 \sim 100$, where the normalized field amplitude
    \begin{equation}
        a_0 \equiv \frac{e \sqrt{- a_\mu a^\mu}}{m} = \frac{e E_L}{ m k_0}
    \end{equation}
    is a Lorentz invariant, $e$ and $m$ are electron charge and mass, $a_\mu$ is the laser $4$-field and $E_L$ is the laser electric field. Laser intensity is anticipated to further reach $10^{23-24}$W/cm$^2$ or even higher on $10-100$PW laser facilities in planning and building, such as ELI \cite{ELI_Website,ELI_Physics01,ELI_Physics02,ELI_Physics03} and XCELS \cite{XCELS}.

    At such intensity, there are two important SFQED processes, nonlinear Compton scattering (NCS)
    \begin{equation}
       e^- (p) + n \gamma_l (k)  \rightarrow \gamma (k') + e^- (p') ,
    \end{equation}
    and its symmetric process, nonlinear Breit-Wheeler process (NBW)
    \begin{equation}
       \gamma (k') + n \gamma_l (k)  \rightarrow e^- (p'') + e^+ (p''') .
    \end{equation}
    Nonlinearity of NCS/NBW stems from high density and coherence of laser photons and is governed by $a_0$. These two highly nonlinear and strongly multi-photon processes are very important both for their high probabilities and dominant roles in future SFQED experiment using ultra intense lasers.

    When laser intensity is not very high ($a_0 \lesssim 1$), radiation formation length of NCS/NBW is of the scale of laser wave length or longer. In this case, the field a NCS/NBW feels is a plane wave or a pulse. As a result, the pulse shape and carrier envelope phase of the pulse can have strong effects in NCS/NBW \cite{Jansen2016,Mackenroth2010,Heinzl2010}. In corresponding theories, NCS/NBW are scatterings of dressed electrons with $\gamma$ photons. Momentum conservation in this case is also replaced by conservation of quasi-momentum
    \begin{equation}
        q^\mu + n k = q'^\mu +k,
    \end{equation}
    where the quasi-momentum
    \begin{equation}
        q^\mu = p^\mu + \frac{e^2 a_0^2}{2(4) k p } k^\mu
    \end{equation}
    for an electron or positron in a circularly (linearly) polarized plane wave. Note that $p$ is the momentum at infinity where the field vanishes. Such theories describe NCS/NBW in lasers with $a_0 \lesssim 1$ well and obtained scattering probabilities can be directly compared to experiments. Scattering angles and deflection angles, which means those between momenta before and after scattering with the whole laser pulse or plane wave at infinity where external field vanishes, can have non-vanishing values in this case. The first experiment of NCS on a laser with $a_0 \lesssim 1$ was carried out in SLAC \cite{SLACExp1996}. Recently, several experiments of NCS in this regime aiming at applications have been reported \cite{Powers2014,Yan2017,Sarri2014}. There were also some researches on NCS in overlapped laser pulses \cite{Hu2016,Wistisen2014}.

    However, in another interesting case of ultra intense lasers ($a_0 \gg 1$), radiation formation length $\delta \phi \sim 1/a_0$ becomes much shorter than the wave length. It means NCS/NBW processes happening between an electron and a laser pulse become isolated local processes. The field a single NCS/NBW feels is also not a plane wave or a pulse but approximately a local constant crossed field that
    \begin{equation}
        F^{\mu\nu}F_{\mu\nu}=\tilde{F}^{\mu\nu} F_{\mu\nu}=0.
    \end{equation}
    Hence the scattering probabilities obtained by above mentioned scattering theories of $e^-$/$e^+$ dressing plane waves or pulses is not that of a single NCS/NBW in this case, but the average over a period or the pulse excluding back reactions and secondary interactions.

    Unfortunately, the condition for strong secondary interaction that $\chi \gtrsim 1$ where
    \begin{equation}
        \chi \equiv e \sqrt{-(F_{\mu\nu}p^\nu)^2} /m^3 \label{ChiDef}
    \end{equation}
    and the condition for strong back reaction that $\alpha a_0 \chi N \gtrsim 1$ \cite{ReviewDiPiazza2012}, where $\alpha = e^2/4\pi \approx 1/137$ is the fine structure constant and $N$ the number of laser cycles, can be easily satisfied in ultra intense lasers whose $a_0 \gg 1$. Therefore back reaction and secondary interactions must be included into consideration when the laser is ultra intense. Otherwise it would be hard to compare theoretical predictions with experiments. Furthermore, the momentum $p$ in the theory is not the instantaneous momentum right before or after a scattering but that at infinity. Hence the simplest way to predict the interaction of an electron with ultra intense lasers is a Monte-Carlo simulation with single NCS/NBW processes in local constant crossed fields described by corresponding differential probabilities.

    Straightforward calculation of NCS/NBW differential probabilities in constant crossed field encounters divergence problems due to temporal symmetry of the field. The clear result of NCS in constant crossed field is the differential probability with respect to
    \begin{eqnarray}
        u                   &=&     \frac{kk'}{kp'}     \nonumber \\
        \tau                &=&     \frac{e \tilde{F}^{\mu \nu} p_\mu p'_\nu}{m^2 a_0 k k'} ,                   .
    \end{eqnarray}
    and the usually adopted one
    \begin{eqnarray}
        \frac{dN_{NCS}}{du dt} & = & F_{\chi,p_0}(u) \nonumber \\
                         & = &  \frac{\alpha}{\pi \sqrt{3}} \frac{m^2}{p_0} \frac{1}{(1+u)^2} \big{[} (1+u+\frac{1}{1+u}) K_{2/3} (\frac{2u}{3\chi})\nonumber \\
                         & \ &   -  \int^{\infty}_{2u/3\chi} dy K_{1/3}(y) \big{]}, \label{NCS_Ritus}
    \end{eqnarray}
    is that with respect to $u$, where $K_\nu(x)$ is the modified Bessel function of $\nu$th order.

    Similarly, NBW differential probability in constant crossed field is
    \begin{eqnarray}
        \frac{dN_{NBW}}{du' dt} & = & G_{\chi', k'_0}(u) \nonumber \\
                         & = &  \frac{\alpha}{\pi \sqrt{3}} \frac{m^2}{k'_0} \big{[} (4u-2) K_{2/3} (\frac{2u}{3\chi'}) \nonumber \\
                         & \ & -  \int^{\infty}_{2u/3\chi'} dy K_{1/3}(y) \big{]}, \label{NBW_Ritus}
    \end{eqnarray}
    where
    \begin{eqnarray}
        u'       & = &    \frac{(kk')^2}{4(kp'')(kp''')}    \nonumber \\
        \chi'   & = &    \frac{e \sqrt{-(F_{\mu\nu} k'^\nu)^2}}{m^3}.   \label{NBW_u}
    \end{eqnarray}
    However, these results do not reflect some very important quantum properties such as the number of photons involved in a single NCS/NBW. Consequently, some basic features such as spectrum, angular distribution and deflection angle of a single NCS/NBW in local constant crossed field of an ultra intense laser are obscure.

    As an alternative, some classical and semi-quantum ideas were introduced into researches of NCS/NBW in ultra intense lasers. In classical electrodynamics, radiation of a relativistic charge with $\gamma=1/\sqrt{1-\beta^2} \gg 1$ concentrates in a cone pointing along instantaneous forward direction with apex angle$\sim 1/\gamma \ll 1$ \cite{JacksonBook,LandauLifshitzBook}. Such narrow angular distribution is usually depicted as forward for simplicity. As a result, radiation reaction on the electron is a recoil pointing backward. The spectrum is also expected to cutoff at the instantaneous energy of electron itself because radiation is usually conceived as a process that a charged particle emits a photon. Since NCS is the major form of electron radiation in ultra intense laser fields, these three classical or semi-quantum ideas on radiation were applied to single NCS in ultra intense lasers. Then $u\approx k'_0/p'_0$ for NCS of an ultra relativistic electron and the energy of emitted $\gamma$ photon and recoiled electron that propagate forward are
    \begin{equation}
        k'_0 \approx \frac{u}{1+u} p_0   \ \ \ \ p'_0 \approx \frac{1}{1+u} p_0.
    \end{equation}
    The spectrum of $\gamma$ photons emitted by NCS given in Eq. (\ref{NCS_Ritus}) then becomes
    \begin{equation}
        \frac{dN_{NCS}}{dt dk'_0} \approx F_{\chi,p_0}(\frac{k'_0}{p_0-k'_0}) \frac{p_0}{(p_0-k'_0)^2}. \label{PopularModelSpectrumFormula}
    \end{equation}
    This understanding of NCS in ultra intense lasers is presently widely accepted and applied \cite{Yan2017,Sarri2014,JiLiangLiang2014,LiJianXin2015,Bell2008,Gonoskov2014,Vranic2014,Blackburn2014,Harvey2017,Green2014,LiJianXin2014,Yu2016,SimAlgorithm01,SimAlgorithm02,SimAlgorithm03,SimAlgorithm04,Simulations01,Simulations02,Simulations03,Gonoskov2015}.

    NBW in ultra intense lasers has a similar model based on similar classical and semi-quantum ideas. In this model, emitted $e^-$ and $e^+$ are along the instantaneous $\gamma$ photon direction. Then $\delta \approx \frac{p''_0}{k'_0}$, $u \approx k_0'^2/(4 p''_0 p'''_0)$ and the spectrum of $e^-$ becomes
    \begin{eqnarray}
        \frac{dN_{NBW}}{dt d \delta} & \approx  & \frac{\alpha m^2}{\sqrt{3} \pi k'_0} \big{[} (\frac{1-\delta}{\delta}+\frac{\delta}{1-\delta}) K_{2/3}(\kappa) \nonumber \\ 
                                     & \ & -  \int_\kappa^\infty K_{1/3} (y) dy \big{]}.
    \end{eqnarray}
    where $\kappa= \frac{2}{3 \chi' \delta (1-\delta)}$.

    These $3$ classical and semi-quantum ideas are the basis of present understanding of single NCS/NBW in ultra intense lasers. Most recent researches concerning NCS/NBW, such as reaction on electron and corresponding applications \cite{Ilderton2013,DiPiazza2008,DiPiazza2009,Harvey2012,JiLiangLiang2014}, its stochasticity \cite{LiJianXin2015,Bell2008,Gonoskov2014,Vranic2014,Blackburn2014,Harvey2017,Green2014,LiJianXin2014,Yu2016} and algorithms to simulate interactions in ultra intense laser \cite{SimAlgorithm01,SimAlgorithm02,SimAlgorithm03,SimAlgorithm04,Simulations01,Simulations02,Simulations03,Gonoskov2015}. Even the few exceptions that did not adopt these ideas, important information of the interaction, such as angular distribution and energy of produced photons, electrons and positrons are avoided by discussing light cone momentum instead \cite{DiPiazza2010,Neitz2013}.

    A simple estimation shows that multi-photon effect will become crucial for single NCS/NBW when laser intensity reaches $I \sim 10^{24}$W/cm$^2$ and present understanding of single NCS/NBW in ultra intense lasers based on the classical and semi-quantum ideas would become invalid. It is well known that when $a_0 \gg 1$, the scattered photon number in NCS/NBW has $n \sim \mathcal{O}(a_0^3)$ \cite{ReviewDiPiazza2012,Ritus1985,Seipt2017,Seipt2013,Ilderton2009}. Considering $k_0 \sim 1$eV in foreseeable future ultra intense lasers, and estimate $p_0$ as its vibration energy $a_0 m$ or present laser wake field acceleration record of a few GeV \cite{Leemans2014}, one gets that $nk_0$ reaches the scale of $p_0$ when $I \sim 10^{24}$W/cm$^2$ ($a_0 \sim 10^3$). Then let us take NCS as example, when the total momentum of scattered laser photons $n\textbf{k}$ becomes comparable to $\textbf{p}$, the total momentum of the scattering system no longer inclines forward. In this case, $\textbf{k}'$ would strongly deviate from instantaneous electron forward direction. It suggests that, on quantum level, multi-photon effects would dominate single NCS/NBW in ultra intense lasers such as future $100$PW facilities.

    In this paper, we first extract the number of laser photons involved in a single NCS/NBW in ultra intense local constant crossed field from corresponding formulae for NCS/NBW in circularly polarized plane wave field in section \ref{sec:number}. Features of multi-photon effects including fixed emission angle, deflection of electron, disappearance of spectrum cutoff for a single NCS in ultra intense lasers are given in section \ref{sec:single}. Non-vanishing emission angle, disappearance of spectrum upper limit and appearance of an spectrum lower limit for NBW are deduced and given in section \ref{sec:single_NBW}. Considering back reaction and secondary interaction of NCS and NBW are very important in future SFQED experiments of NCS/NBW, head-on collision of an electron beam with an ultra intense laser pulse is simulated in section \ref{sec:succs} to show multi-photon effects in NCS/NBW in future experiments. Finally, conclusions are given in section \ref{sec:concl}.

\section{Strong Field Approximation of Scattered Laser Photon Number in a Single Nonlinear Compton Scattering/Nonlinear Breit-Wheeler Process}\label{sec:number}
    It is a well known conclusion of SFQED that the number $n$ of laser photons an electron scatters in a single NCS process when it interacts with an ultra intense laser ($a_0 \gg 1$) is of the order of $a_0^3$ \cite{Ritus1985,ReviewDiPiazza2012,Seipt2013}. However, to investigate the effect of these laser photons in a single NCS, the qualitative relation $n\sim a_0^3$ is far from enough, a quantitative differential probability $W$ with respect to $n$ is needed.

    As discussed in section \ref{sec:intro}, radiation formation length $\delta \phi \sim 1/a_0$ of NCS \cite{Ritus1985,ReviewDiPiazza2012} is much shorter than one period in ultra intense lasers with $a_0 \gg 1$. In this case, the field a single NCS feels is approximately a constant crossed field rather than a plane wave field. Hence it is the differential probability $\partial W^{CF}/\partial n$ of NCS in constant crossed field rather than other conceptions such as quasi scattered photon number in circularly polarized plane wave field that determines NCS (superscript $CF$ stands for constant crossed field) in ultra intense lasers.

    The ordinary method to get $\partial W^{CF}/\partial n$ encounters divergence problems introduced by temporal symmetry of constant crossed fields \cite{Ritus1985}. To avoid this, we will try to extract $\partial W^{CF}/\partial n$ from $W^{CP}$, the NCS differential probability of an electron with circularly polarized ($CP$) plane wave field \cite{Ritus1985,BerestetskiiBook,Ilderton2009} excluding back reaction and secondary interactions, which is free from divergence.

    A physically correct differential probability $W$ of NCS should be Lorentz and gauge invariant. Hence it is a function of corresponding invariant parameters. The initial state of NCS includes $3$ Lorentz vectors, $a^\mu$, $k^\mu$ and $p^\mu$. Their combinations give $4$ independent non vanishing parameters:
    \begin{eqnarray}
        \chi                &=&     \frac{e \sqrt{- (F^{\mu\nu} p_\nu)^2 } }{m^3}   \nonumber \\
        a_0                 &=&     \frac{e \sqrt{- a_\mu a^\mu}}{m}                \nonumber \\
        \textit{f}          &=&     \frac{e^2 F^{\mu\nu} F_{\mu\nu}}{4m^4}          \nonumber \\
        \textit{g}          &=&     \frac{e^2 F^{\mu\nu} \tilde{F}_{\mu\nu}}{4m^4}.
    \end{eqnarray}
    Here, $e$, $m$ and $p$ are charge, mass and $4$-momentum of electron, $F$ and $a$ are tensor and $4$-vector of the field. Among these $4$ parameters, the last two vanish for constant crossed field therefore $\chi$ and $a_0$ distinguish physically different initial states of NCS in constant crossed fields that cannot be connected by Lorentz transformation. Thus they determine $W^{CF}$, i. e., $W^{CF}=W^{CF}_{\chi,a_0}$.

    A set of independent Lorentz and gauge invariant parameters for final state includes the scattered photon number $n$ and
    \begin{eqnarray}
        u                   &=&     \frac{kk'}{kp'}     \nonumber \\
        \tau                &=&     \frac{e \tilde{F}^{\mu \nu} p_\mu p'_\nu}{m^2 a_0 k k'}                       .
    \end{eqnarray}
    Hence the differential probability is a function of $n$, $u$ and $\tau$, i. e., $W^{CF}_{\chi,a_0}(n,u,\tau)$. Note that although $n$ is the number of scattered laser photon number, it is not predetermined before scattering therefore is a parameter for final state .

    Consider the NCS in a linearly polarized plane wave field concentrates around $|\tau |\lesssim 1$ when $a_0\gg 1$, the deviation of radiated photon from the $a-p$ plane $\sim \tau/a_0 \rightarrow 0$ is strongly suppressed in the frames where $k$ and $p$ are antiparallel \cite{Ritus1985}. Hence the differential probability is approximately $W^{CF}_{\chi,a_0}(n,u)\delta(\tau)$.

    In $a_0\gg 1$ limit, NCS of an electron with $4$-momentum $p^\mu_0$ at infinity with a circularly polarized laser field is approximately the average of this electron with instantaneous $p^\mu=p^\mu (\phi)$ in a constant crossed field $F^{CF}=F^{CP}(\phi)$ over a period without considering radiation reaction, i. e.,
    \begin{equation}
        W^{CP}_{\chi, a_0}(n_{CP},u) \approx  \frac{1}{2\pi} \int_0^{2\pi} d\phi W^{CF}_{\chi(\phi),a_0(\phi)} (n(n_{CP}),u(\phi)), \label{RelCFCP_01}
    \end{equation}
    where $\chi$ and $u$ are defined with electron momentum at infinity $p^\mu_0$. $\chi(\phi)$ and $u(\phi)$ are instantaneous values given by instantaneous electron momentum $p^\mu(\phi)$ which is
    \begin{eqnarray}
        p_\perp(\phi)          &=& e a(\phi)+p_{\perp 0}   \nonumber \\
        p^0(\phi) - p_{\parallel}(\phi)    &=& p^0_0 - p_{\parallel 0} = \alpha \nonumber \\
        p_{\parallel}(\phi)    &=& \frac{m^2 - \alpha^2 + p_\perp(\phi)^2}{2\alpha} \nonumber \\
        p^0(\phi)    &=& \frac{m^2 + \alpha^2 + p_\perp(\phi)^2}{2\alpha}
    \end{eqnarray}
    where $p_\parallel$ is the projection of $\textbf{p}$ along $\textbf{k}$ direction and $p_\perp= p - p_\parallel$. $n(n^{CP})$ reflects the difference between the instantaneous scattered photon number $n$ and the $n^{CP}$ defined in quasi-momentum conservation as
    \begin{equation}
        q^\mu+n_{CP} k^\mu= q'^\mu + k'^\mu,
    \end{equation}
    where
    \begin{eqnarray}
        q^\mu  &=& p^\mu_0  + \frac{m^2 a_0^2}{2 kp} k^\mu \nonumber \\
        q'^\mu &=& p'^\mu_0 + \frac{m^2 a_0^2}{2 kp'} k^\mu \label{Defq}
    \end{eqnarray}
    are quasi-momenta of the electron before and after scattering in circularly polarized plane wave field \cite{Ritus1985}.

    Considering the circular polarization and $p(\phi) k = p_0 k$, $a_0(\phi)=a_0$ is a constant, $\chi(\phi)=\chi$ and $u(\phi)=u$. In other words, the differential probability $W^{CF}_{\chi a_0}$ for
    \begin{eqnarray}
        e (p^\mu(\phi)) + n \gamma_L (k) \rightarrow e (p'^\mu (\phi)) + \gamma (k'^\mu)
    \end{eqnarray}
    in the constant crossed field $F^{CF}=F(\phi)$ is the same to that for
    \begin{eqnarray}
        e (p^\mu_0) + n \gamma_L (k) \rightarrow e (p'^\mu_0) + \gamma (k'^\mu)
    \end{eqnarray}
    in the same $F^{CF}(\phi)$. Then Eq. (\ref{RelCFCP_01}) degenerates to
    \begin{equation}
        W^{CP}_{\chi, a_0}(n_{CP},u) \approx  \frac{1}{2\pi} \int_0^{2\pi} d\phi W^{CF}_{\chi,a_0} (n(n_{CP}),u). \label{RelCFCP_02}
    \end{equation}

    Combined with the definition of quasi-momenta in Eq. (\ref{Defq}), the relation between $n$ and $n^{CP}$ is
    \begin{equation}
        n=n^{CP} -  \frac{u}{2\chi} a_0^3,    \label{NCNP}
    \end{equation}
    and Eq. (\ref{RelCFCP_02}) becomes
    \begin{equation}
        W^{CF}_{\chi,a_0} (n,u) \approx W^{CP}_{\chi, a_0}(n+\frac{u}{2\chi} a_0^3,u). \label{RelCFCP_04}
    \end{equation}

    The differential probability $W^{CP}$ was given in Ref. \cite{Ritus1985,BerestetskiiBook,Ilderton2009} as
    \begin{eqnarray}
        & \ & W^{CP}_{\chi,a_0} (u, n) \nonumber \\
        &=&  \frac{e^2 m^2}{4q_0} \frac{1}{(1+u)^2} [-4 J_n^2(z_n(u))+ a_0^2 (\frac{u^2+2u+2}{1+u})  \nonumber \\
              & \ &(J^2_{n+1} (z_n)+J^2_{n-1}(z_n)-2J^2_n((z_n))], \label{WC2}
    \end{eqnarray}
    where
    \begin{eqnarray}
        y_n     &=&     \frac{2n kp}{m_*^2}                                             \nonumber \\
        m_*^2   &=&     m^2 (1+a_0^2)                                                   \nonumber \\
        z_n(u)  &=&     \frac{2a_0}{y_1} \sqrt{\frac{u (y_n-u)}{1+a_0^2}}.
    \end{eqnarray}

    Eq. (\ref{RelCFCP_04}) together with Eq. (\ref{WC2}) are still hard for direct application because it includes complex combinations of $J$, some important features such as $n \sim a_0^3$ is also obscure. To overcome these problems, approximations and simplifications are needed.

    In the $J_n(z)$ appeared in Eq. (\ref{WC2}), $z$ distributes from $0$ to $ n a_0/\sqrt{1+a_0^2} \approx n$. In this region, $J_n(z)$ in large $n$ limit is highly suppressed except when $z \approx n$. Since both terms of Eq. (\ref{WC2}) are only significant when $n-z  \lesssim n^{1/3}$, the physically important part of $W^{CP}$ comes from regions where $u \approx y_n/2 $, or more specifically,
    \begin{equation}
        |u-y_n/2| \lesssim \frac{y_n}{n^{1/3}}.
    \end{equation}
    In this region,
    \begin{equation}
        \frac{u}{2\chi} a_0^3 = \frac{n^{CP}}{2} (1- a_0^{-2} + \mathcal{O}(n^{-1/3}))
    \end{equation}
    and
    \begin{equation}
        n^{CF}=\frac{n^{CP}}{2}(1 + a_0^{-2}+\mathcal{O}(n^{-1/3}).
    \end{equation}

    Hence when $a_0 \gg 1$, NCS in constant crossed field is significant only when
    \begin{equation}
        n^{CF} \approx \frac{u}{2\chi} a_0^3. \label{nCCF}
    \end{equation}
    Since the \textquotedblleft $u$ spectrum\textquotedblright \ $F_{\chi,a_0}(u)$ in constant crossed fields is given in Eq. (\ref{NCS_Ritus}), we can arrive at the strong field approximation for NCS differential probability
    \begin{equation}
        W^{CF}_{\chi,a_0}(n,u) \approx F_{\chi,a_0}(u) \delta(n-\frac{u}{2\chi} a_0^3). \label{uSpectrum_NPSIncl}
    \end{equation}
    Note the relation $n\sim \mathcal{O}(a_0^3)$ is also recovered since $u/2\chi$ is of the scale of $1$.

    Since NBW is the symmetric process of NCS, deduction for the involved laser photon number in a single NBW in ultra intense lasers is quite similar, which gives
    \begin{equation}
        n_{NBW} \approx \frac{2 u'}{ \chi'} a_0^3   \label{nCCF_NBW}
    \end{equation}
    when $a_0 \gg 1$, where the parameter $u'$ of NBW is given in Eq. (\ref{NBW_u}).

\section{Multi-Photon Effects in a Single Nonlinear Compton Scattering in Ultra Intense Lasers}\label{sec:single}
    Then we apply the strong field approximation for NCS differential probability of in Eq. (\ref{uSpectrum_NPSIncl}) to investigate multi-photon effects in NCS.

    In frames where $\textbf{k}$ and $\textbf{p}$ are antiparallel, once $u$ and $n$ are fixed, consider NCS is concentrated around the $\textbf{a}$-$\textbf{p}$ plane \cite{Ritus1985,Seipt2013}, possible choice of $\textbf{k}'$ decreases to two vectors. One of them is on the $\textbf{k}-\textbf{a}-\textbf{p}$ half plane and one on the other half.

    These two vectors are distinguished by the Lorentz gauge invariant
    \begin{equation}
        \rho = \frac{e F^{\mu\nu} p'_\mu p_\nu }{m^2 a_0^2 k k'}
    \end{equation}
    In this specific frame, it becomes $e \vec{k'}\cdot \vec{ a} kp/ a_0^2 m^2 kk' $, hence the vector on $\textbf{p}-\textbf{a}-\textbf{k}$ half-plane corresponds to $\rho>0$ and the other has $\rho<0$.

    \begin{figure}
        \begin{center}
            \includegraphics[width=0.40\textwidth]{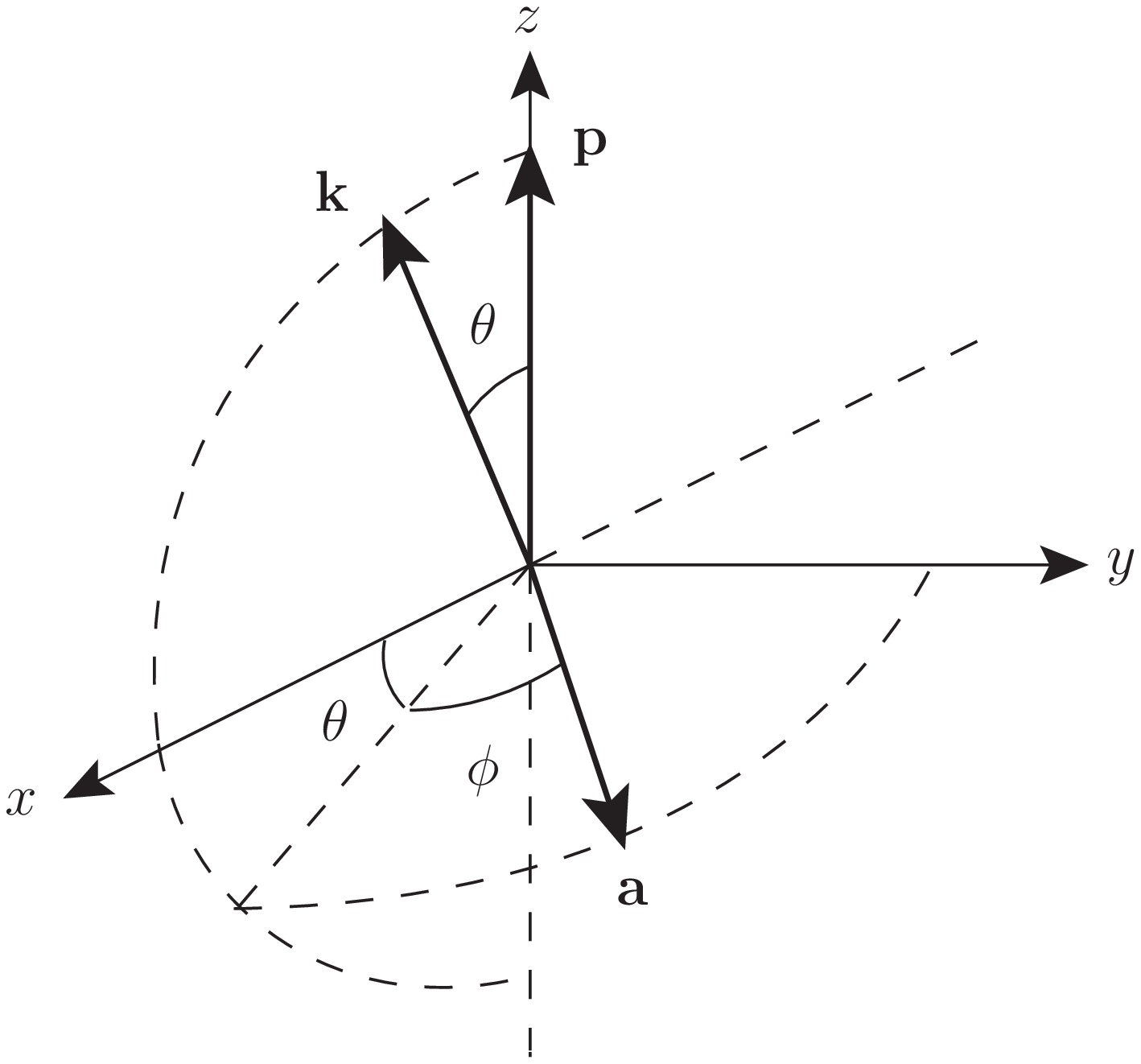}
        \end{center}
        \caption{$\textbf{p}$, $\textbf{k}$ and $\textbf{a}$ in a general frame.}
        \label{Frames}
    \end{figure}

    Then we will discuss the emitted $\gamma$ photon energy $k'_0$ and direction $\hat{e}_{\textbf{k}'}$ of NCS of ultra relativistic electrons. As mentioned above, in large $a_0$ limit, physical $k'$ has $\tau\approx 0$. In cases where $\textbf{k}$ and $\textbf{p}$ are parallel or anti-parallel, this is just the $\textbf{a}-\textbf{p}$ plane. Solve
    \begin{equation}
        \left\{
        \begin{aligned}
            nkp' &= k'p  \\
            u    &= \frac{kk'}{kp'} \\
            n    &= \frac{u}{2\chi}a_0^3
        \end{aligned}
        \right. ,
    \end{equation}
    one can get the radiated $\gamma$ photon energy and the emission angle are
    \begin{eqnarray}
        k'_0                &\approx & \frac{u (1+C)}{1+u} p_0   \nonumber \\
        \cos \theta_{k'}^A    &\approx & \frac{1-C}{1+C}           \label{EmiParaA}
    \end{eqnarray}
    where $C=a_0^3 k_0/2\chi p_0$, $\theta$ is the angle between $\textbf{k}'$ and $\textbf{p}$ and $A$ stands for that $\textbf{k}$ and $\textbf{p}$ are antiparallel. Apparently, these two solutions for $\textbf{k}'$ distribute symmetrically with respect to $\textbf{p}$.

\begin{figure}
        \begin{center}
            \includegraphics[width=0.23\textwidth]{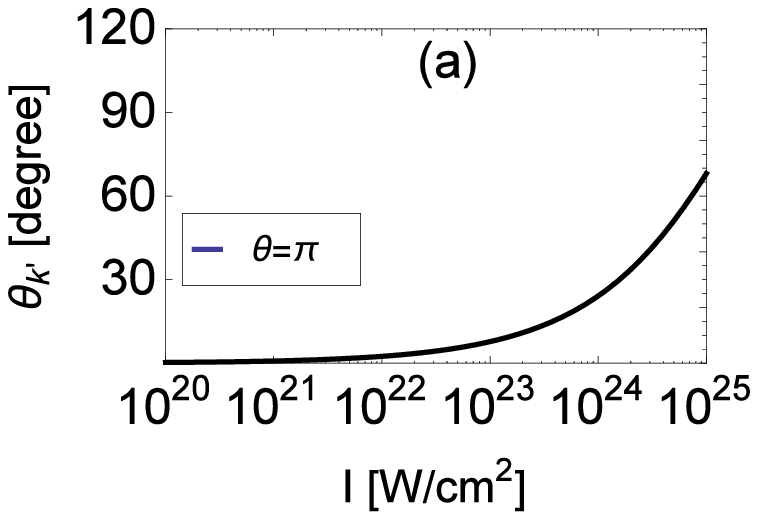}
            \includegraphics[width=0.23\textwidth]{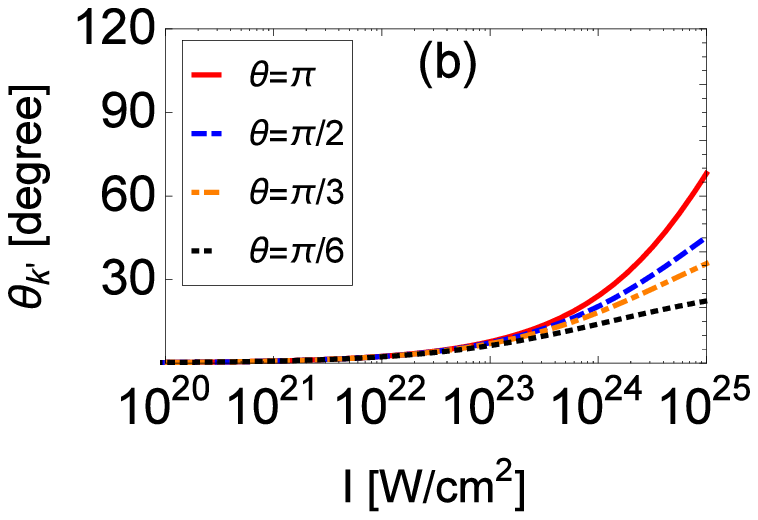} \\
            \includegraphics[width=0.23\textwidth]{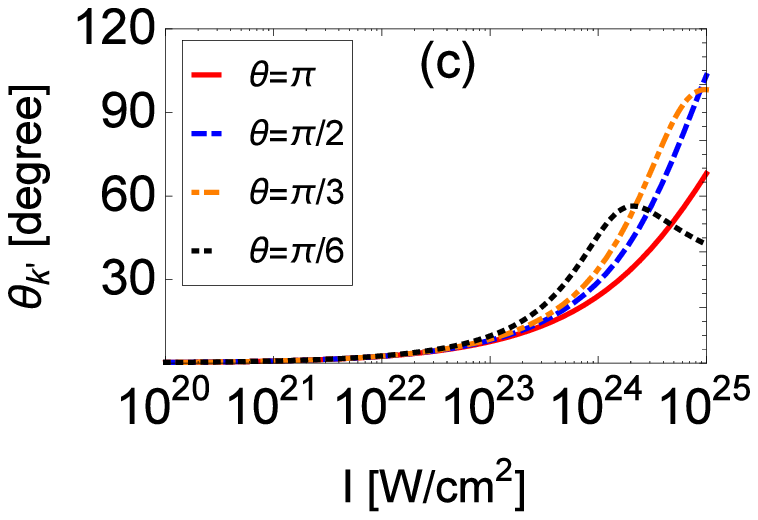}
            \includegraphics[width=0.23\textwidth]{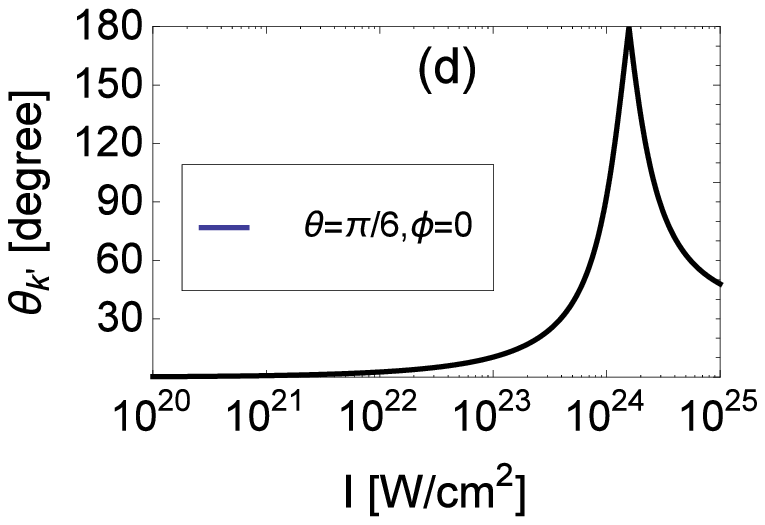}
        \end{center}
        \caption{Dependence of instantaneous emission angle on intensity, $\theta_{k'\pm}$ when $\textbf{k}$ and $\textbf{p}$ are antiparallel (a), $\theta_{k'+}$ when $\phi=\pi/6$ (b),  $\theta_{k'-}$ when $\phi=\pi/6$ (c) and $\theta_{k'-}$ when $\theta=\pi/6$ and $\phi=0$ (d). Corresponding parameters are $p_0=1.022$GeV, $k_0=1.24$eV ($\lambda=1\mu$m) and $I=a_0^2 1.37\times 10^{18}$W/cm$^2$.}
        \label{EmissionAngle}
    \end{figure}

    On the other hand, energy and deflection angle of the electron after a single NCS are
    \begin{eqnarray}
        p'_0                &\approx &  \frac{1+u^2 C}{1+u} p_0 \nonumber\\
        \cos \theta_{p'}^A    &\approx &  \frac{1-u^2 C}{1+u^2C}  \label{DeflParaA}
    \end{eqnarray}
    Note that $\theta_{k'}^A$ is independent of $u$ while $\theta_{p'}^A$ is not, which means the emission angle is fixed while the deflection angle is not. Fig. \ref{EmissionAngle} (a) shows the emission angle of an $1$GeV electron in such head-on case, it reaches $\sim 30^\circ$ at $I=10^{24}$W/cm$^2$.

    Results obtained in this frame can be extended to arbitrary frames through Lorentz transformation. Without loss of generality, we fix $\textbf{p}$
    on the $z$ axes, $\textbf{k}$ on $x-z$ plane as shown in Fig. \ref{Frames}. The angle between $\textbf{p}$ and $\textbf{k}$ is $\theta$ and the angle between $\textbf{a}$ and $\textbf{p}-\textbf{k}$ plane is $\phi$.

    Considering the ultra relativistic condition, a boost along the $(\hat{e}_\textbf{k}+\hat{e}_\textbf{p})/2$ direction with $\beta=\cos \frac{\theta}{2}$ transforms $\textbf{k}$ and $\textbf{p}$ antiparallel. In the boosted frame, Eq. (\ref{EmiParaA}) is applicable, therefore
    \begin{eqnarray}
        k^{'A}_0             & = & \frac{u (1+C)}{1+u} p_{0}^A   \nonumber \\
        \cos \theta_{k'}^A   & = & \frac{1-C}{1+C},
    \end{eqnarray}
    where the superscript $A$ stands for boosted anti-parallel frame and $p_0^A=\sin \frac{\theta}{2} p_0$.

    Apply the reverse lorentz boost, $k'_{\pm}$ of emitted $\gamma$ photon in the laboratory frame is
    \begin{eqnarray}
        \frac{u (1+C) p_0}{1+u}\begin{pmatrix}  1 \mp \cos\phi\cos\frac{\theta}{2}\sin\theta_{k'}^A \\
                                        \sin\theta \sin^2\frac{\theta_{k'}^A}{2} \mp \sin\frac{\theta}{2} \cos\phi\sin\theta_{k'}^A \\
                                        \pm \sin\frac{\theta}{2} \sin\theta_{k'}^A\sin\phi \\
                                        1- 2\sin^2\frac{\theta_{k'}^A}{2}\sin^2 \frac{\theta}{2} \mp \cos\phi\cos\frac{\theta}{2}\sin\theta_{k'}^A
                                    \end{pmatrix}.  \nonumber \\ \ \label{GeneralkP1}
    \end{eqnarray}
    where $\pm$ is the sign of $\rho$. Corresponding energy and emission angle with respect to $\textbf{p}$ are
    \begin{eqnarray}
        \cos \theta_{k'\pm}   &=& \frac{1- 2\sin^2\frac{\theta_{k'}^A}{2}\sin^2 \frac{\theta}{2}  \mp \cos\phi\cos\frac{\theta}{2} \sin\theta_{k'}^A }{1 \mp \cos \phi \cos \frac{\theta}{2} \sin \theta_{k'}^A } \nonumber \\
        k'_{0,\pm}            &=& \frac{u(1+C)}{1+u} p_0 (1 \mp \cos \phi \cos \frac{\theta}{2} \sin \theta_{k'}^A).  \label{GeneralkP2}
    \end{eqnarray}

    Note that Eq. (\ref{GeneralkP1}) and Eq. (\ref{GeneralkP2}) degenerate to the antiparallel case results in Eq. (\ref{EmiParaA}) when $\theta=\pi$, and the two $\textbf{k}'$ are symmetric with respect to $\textbf{p}$ when $\phi=\pi/2$ or $\theta=\pi$. Fig. \ref{EmissionAngle} (b) and (c) show emission angles $\theta_{k'+}$ and $\theta_{k'-}$ of a $1$GeV electron in intense fields when $\phi$ is fixed at $\pi/6$. Fig. \ref{EmissionAngle} (d) gives the emission angle when $\textbf{k}$, $\textbf{p}$ and $\textbf{a}$ are on the same plane. It shows that when $\phi$ is small, emission angle can be very large, even backward around $10^{24}$W/cm$^2$.

    \begin{figure}
        \begin{center}
            \hspace{-0.5cm}
            \includegraphics[width=0.23\textwidth]{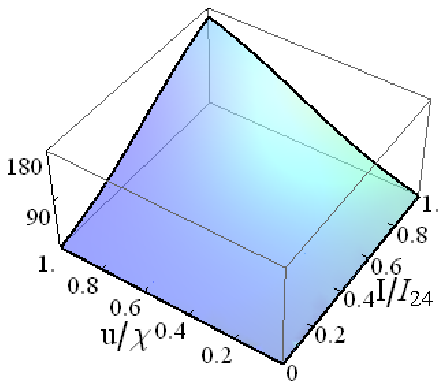}
            \includegraphics[width=0.23\textwidth]{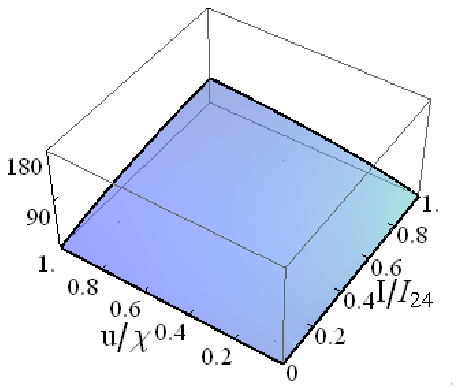}
            \hspace{-0.5cm}
        \end{center}
        \caption{Instantaneous deflection angles $\theta_{p'-}$ (left) and $\theta_{p'+}$ (right) in degree, $u$ is normalized by $\chi$ and $I$ by $10^{24}$W/cm$^2$. Corresponding parameters are $\gamma=2000$, $\theta=\pi/2$ and $\phi=0$.}
        \label{DefAngle}
    \end{figure}

    The momentum $\textbf{p}'_{\pm}$ of deflected electron in general case and laboratory frame is approximately
    \begin{eqnarray}
        \frac{(1+u^2 C) p_0}{1+u}\begin{pmatrix}
                                        \sin\theta \sin^2\frac{\theta_{p'}^A}{2} \pm \sin\frac{\theta}{2} \cos\phi\sin\theta_{p'}^A \\
                                        \mp \sin\frac{\theta}{2} \sin\theta_{p'}^A\sin\phi \\
                                        1- 2\sin^2\frac{\theta_{p'}^A}{2}\sin^2 \frac{\theta}{2} \pm \cos\phi\cos\frac{\theta}{2}\sin\theta_{p'}^A
                                    \end{pmatrix}. \nonumber \\ \ \label{GeneralpP1}
    \end{eqnarray}
    Corresponding energy and deflection angle with respect to the electron instantaneous forward direction are
    \begin{eqnarray}
        \cos \theta_{p'\pm}   &=& \frac{1- 2\sin^2\frac{\theta_{p'}^A}{2}\sin^2 \frac{\theta}{2}  \pm \cos\phi\cos\frac{\theta}{2} \sin\theta_{p'}^A }{1 \pm \cos \phi \cos \frac{\theta}{2} \sin \theta_{p'}^A }\nonumber \\
        p'_{0,\pm}           &=& \frac{1+u^2 C}{1+u} p_0 (1 \pm \cos \phi \cos \frac{\theta}{2} \sin \theta_{p'}^A). \label{GeneralpP2}
    \end{eqnarray}
    Fig. \ref{DefAngle} shows deflection angle $\theta_{p'\pm}$ of an $1$GeV electron when $\theta=\pi/2$ and $\phi=0$. Parameter $u$ is normalized by $\chi$ since the probability of $u>\chi$ is very small. The deflection angle grows fast with laser intensity, when $I$ surpasses $10^{23}$W/cm$^2$, deflection angle gets apparently non vanishing value. At $10^{24}$W/cm$^2$, a considerable part of $\theta_{p'-}$ is even larger than $\pi/2$.

    \begin{figure}
        \begin{center}
            \includegraphics[width=0.23\textwidth]{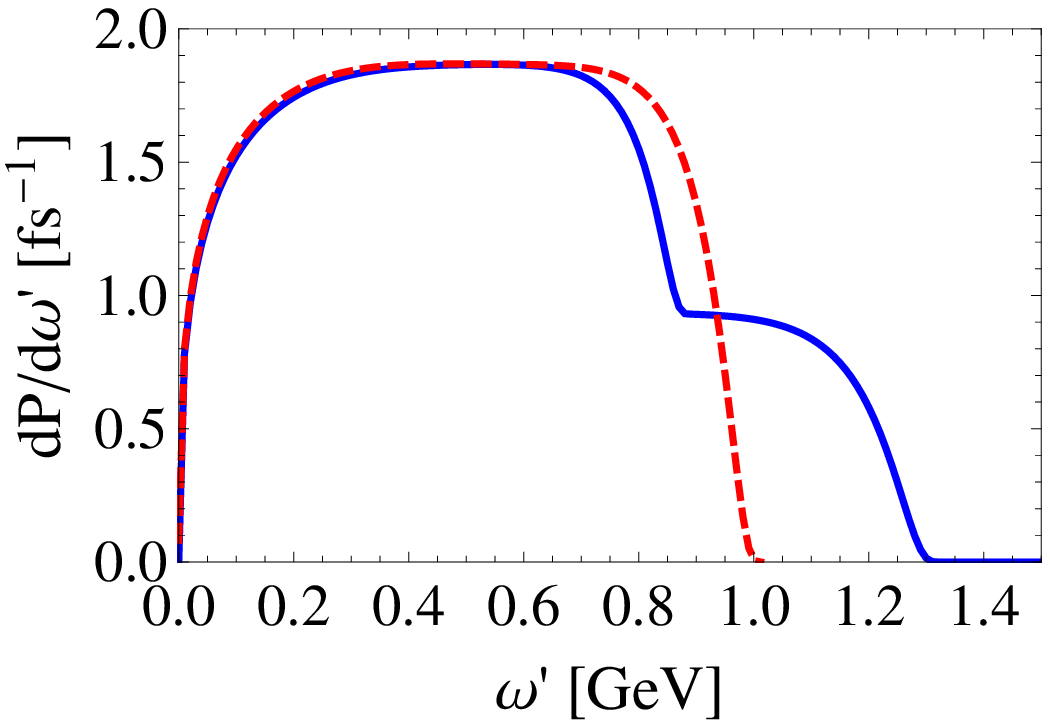}
            \includegraphics[width=0.235\textwidth]{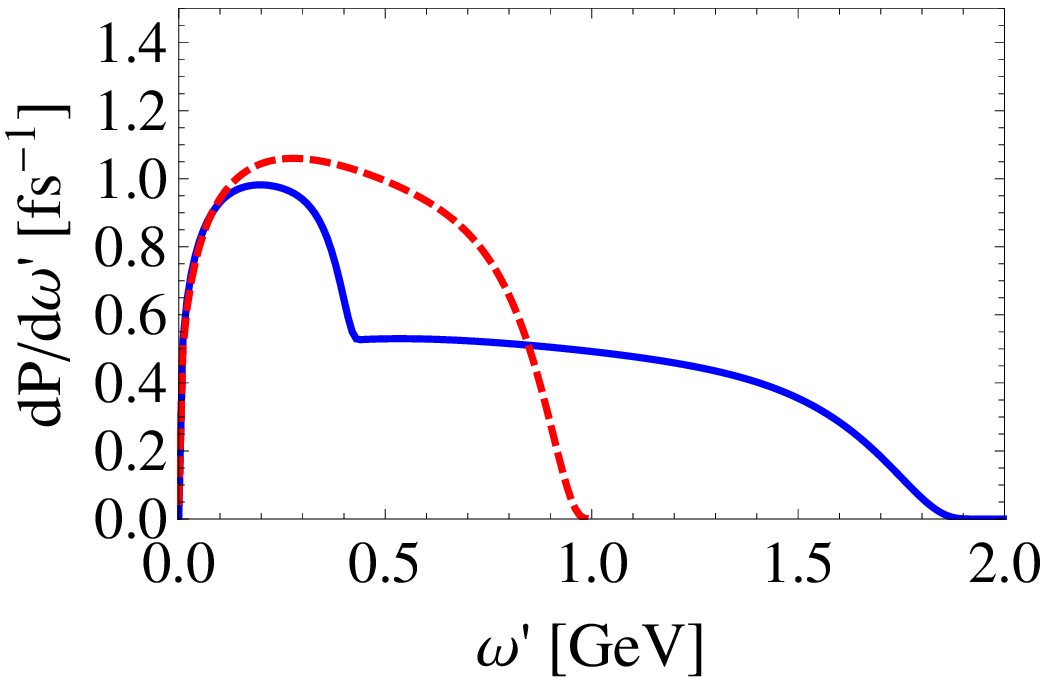}
            \vspace{-0.5cm}
        \end{center}
        \caption{Instantaneous energy integrated spectrum of a $1$GeV electron including (blue solid) and excluding (red dashed) multi-photon effects. Field intensity is $10^{24}$W/cm$^2$, $\theta=\pi/2$, $\phi=\pi/3$ for the left panel and $\theta=\pi/3$, $\phi=0$ for the right.}
        \label{Spectrum_Ins}
    \end{figure}

    Spectrum is also very important for NCS. Comparison of energy integrated spectra $dP/d\omega'dt=\omega' dN/d\omega'dt$ of a $1$GeV electron including and excluding multi-photon effects in ultra intense field of $I=10^{24}$W/cm$^2$ is shown in Fig. \ref{Spectrum_Ins}. As given in Eq. (\ref{GeneralkP2}), multi-photon effect blue shifts half of the spectrum by a factor of $(1+C)(1+\cos\phi \cos \frac{\theta}{2} \sin \theta_{p'}^A)$ while red shifts the other half by $(1+C)(1-\cos\phi \cos \frac{\theta}{2} \sin \theta_{p'}^A)$. Hence the spectrum gets a two-stage structure. The blue shifted part can surpass the instantaneous electron energy $p_0=1$GeV, which is the cutoff excluding multi-photon effects, as long as $\phi$ is far from $\pi$ and $\theta$ is not very large.

\section{Multi-Photon Effects in a Single Nonlinear Breit-Wheeler Process in Ultra Intense Lasers}\label{sec:single_NBW}
    To deduce multi-photon effects in a single NBW in ultra intense lasers, we apply the strong field approximation for $n$ in NBW given by Eq. (\ref{nCCF_NBW}).

    In frames where $\textbf{k}$ and $\textbf{k}'$ are antiparallel, once $u$ and $n$ are fixed, consider NBW is concentrated around the $\textbf{a}$-$\textbf{k}'$ plane \cite{Ritus1985}, possible choice of $\textbf{p}''$ decreases to two vectors. One of them is on the $\textbf{k}-\textbf{a}-\textbf{k}'$ half plane and one on the other half plane.

    These two vectors are distinguished by $\rho$, which is $- e \vec{k'} \cdot \vec{ a} kp'''/ a_0^2 m^2 kk' $ in this specific frame. The vector on $\textbf{k}'-\textbf{a}-\textbf{k}$ half-plane corresponds to $\rho<0$ and the other has $\rho>0$.

    \begin{figure}
        \begin{center}
            \includegraphics[width=0.40\textwidth]{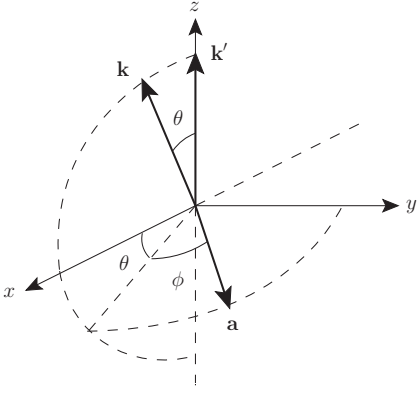}
        \end{center}
        \caption{Geometry of $\textbf{k}'$, $\textbf{k}$ and $\textbf{a}$ of a NBW process in arbitrary frame.}
        \label{Frames_NBW}
    \end{figure}

    In the comparatively easy case that $\textbf{k}$ and $\textbf{k}'$ are anti-parallel, the energy of emitted $e^-$ and $e^+$ are
    \begin{eqnarray}
        p''^A_0   & = &   \frac{k'_0+n k_0}{2}\big{[} 1+ (2\delta -1) \zeta \big{]}       \nonumber \\
        p'''^A_0  & = &   \frac{k'_0+n k_0}{2}\big{[} 1- (2\delta -1) \zeta \big{]}.
    \end{eqnarray}
    Their angles to $\textbf{k}'$ are
    \begin{eqnarray}
        \cos \theta^A_{p''}     & = &   \frac{\zeta+(2\delta-1)}{1+(2\delta-1) \zeta}   \nonumber \\
        \cos \theta^A_{p'''}    & = &   \frac{\zeta-(2\delta-1)}{1-(2\delta-1) \zeta},
    \end{eqnarray}
    where $\delta=kp''/kk'$,
    \begin{equation}
        \zeta = \frac{k'_0  -n k_0}{k'_0 + nk_0},
    \end{equation}
    and $n$ is the involved laser photon number in a single NBW in ultra intense lasers given in Eq. (\ref{nCCF_NBW}).

    Similar to NCS, when $a_0 \gg 1$, NBW also has analytic expression in the general case. As shown in Fig. \ref{Frames_NBW}, without loss of generality, $\textbf{k}'$ is fixed along the positive z direction and $\textbf{k}$ on the $z-x$ plane. In this frame, momentum $\textbf{p}''$ of produced $e^-$ including multi-photon effect is
    \begin{eqnarray}
        \textbf{p}''_{\pm} & = & \frac{(k'_0+nk_0)}{2}(1+(2\delta-1)\zeta) \nonumber \\ & \ & \begin{pmatrix}
                                                            \sin \theta \sin^2\frac{\theta_{p''}^A}{2} \mp \sin \frac{\theta}{2} \cos\phi\sin\theta_{p''}^A \\
                                                            \pm  \sin \frac{\theta}{2}  \sin\theta_{p''}^A \sin\phi \\
                                                            1- 2 \sin^2 \frac{\theta_{p''}^A}{2} \sin^2 \frac{\theta}{2}  \mp \cos \frac{\theta}{2} \cos\phi \sin\theta_{p''}^A
                                                            \end{pmatrix},    \nonumber \\ \ \label{GeneralpPP_NBW}
    \end{eqnarray}
    Corresponding energy and emission angle with respect to the $\gamma$ photon forward direction are
    \begin{equation}
        p''^0_{\pm}              =    \frac{k'_0+nk_0}{2} [1+ (2\delta-1) \zeta] (1 \mp \cos \phi \cos \frac{\theta}{2} \sin \theta^A_{p''} ) 
    \end{equation}
    and
    \begin{equation}
        \cos \theta_{p'' \pm}    =    \frac{1- 2 \sin^2 \frac{\theta}{2} \sin^2 \frac{\theta^A_{p''}}{2} \mp \cos\phi \cos\frac{\theta}{2} \sin \theta^A_{p''}  }{1 \mp \cos \phi \cos \frac{\theta}{2} \sin \theta^A_{p''}}
    \end{equation}

    The momentum $\textbf{p}'''$ of produced $e^+$ is then
    \begin{eqnarray}
        \textbf{p}'''_{\pm} &=&  \frac{(k'_0+nk_0)}{2}(1-(2\delta-1)\zeta)\nonumber \\ & \ & \begin{pmatrix}
                                                            \sin \theta \sin^2\frac{\theta_{p'''}^A}{2} \pm  \sin \frac{\theta}{2} \cos\phi\sin\theta_{p'''}^A \\
                                                            \mp \sin \frac{\theta}{2} \sin\theta_{p'''}^A \sin\phi \\
                                                            1-  2 \sin^2 \frac{\theta}{2} \sin^2 \frac{\theta_{p'''}^A}{2}  \pm \cos\phi \cos\frac{\theta}{2}\sin\theta_{p'''}^A
                                                            \end{pmatrix}, \nonumber \\ \ \label{GeneralpPPP_NBW}
    \end{eqnarray}
    Corresponding energy and emission angle with respect to the $\gamma$ photon forward direction are
    \begin{equation}
        p'''^0_{\pm}              =    \frac{k'_0+nk_0}{2} [1- (2\delta-1) \zeta] (1 \pm \cos \phi \cos \frac{\theta}{2} \sin \theta^A_{p'''} )  
    \end{equation}
    and
    \begin{equation}
        \cos \theta_{p''' \pm}    =    \frac{1- 2 \sin^2 \frac{\theta}{2} \sin^2 \frac{\theta^A_{p'''}}{2} \pm \cos\phi \cos\frac{\theta}{2} \sin \theta^A_{p'''}  }{1 \pm \cos \phi \cos \frac{\theta}{2} \sin \theta^A_{p'''}}.
    \end{equation}

    Multi-photon effects on a single NBW in ultra intense lasers is similar but different to that on NCS. As discussed at the end of section \ref{sec:single}, in a single NCS in ultra intense fields, it shifts half of the emitted $\gamma$ spectrum by a factor of $(1+C)(1 +\cos\phi \cos \frac{\theta}{2} \sin \theta_{p'}^A)$ and the other half by $(1+C)(1 - \cos\phi \cos \frac{\theta}{2} \sin \theta_{p'}^A)$. Hence the cutoff of spectrum is shifted by a factor of $(1+C)(1 + |\cos\phi \cos \frac{\theta}{2} \sin \theta_{p'}^A)|$.

\begin{figure}
        \begin{center}
            \includegraphics[width=0.23\textwidth]{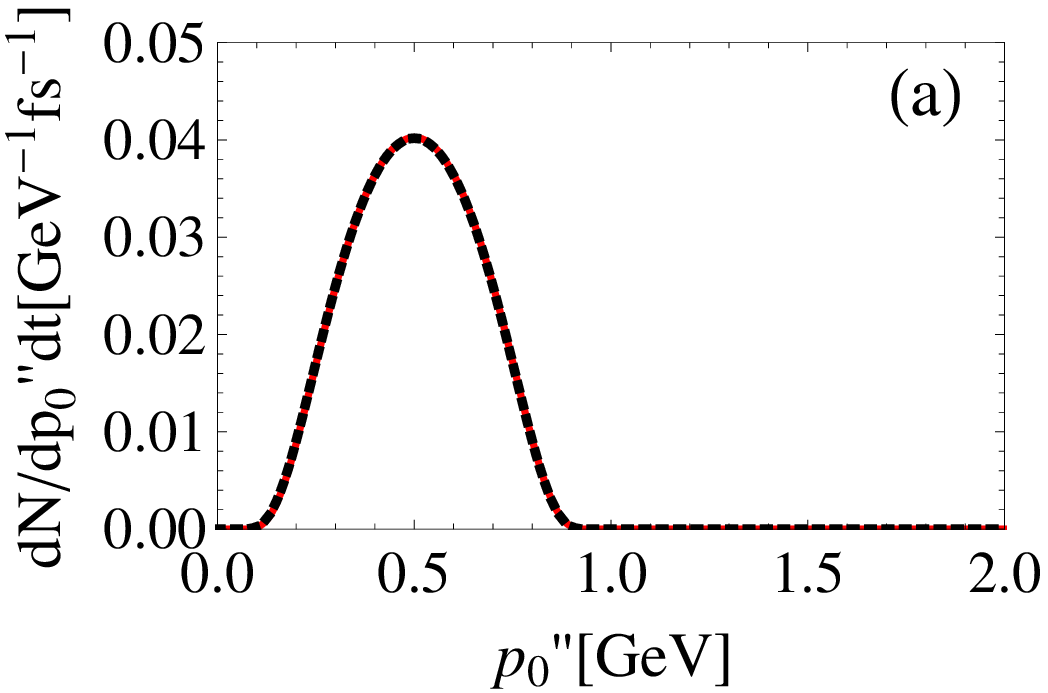}
            \includegraphics[width=0.23\textwidth]{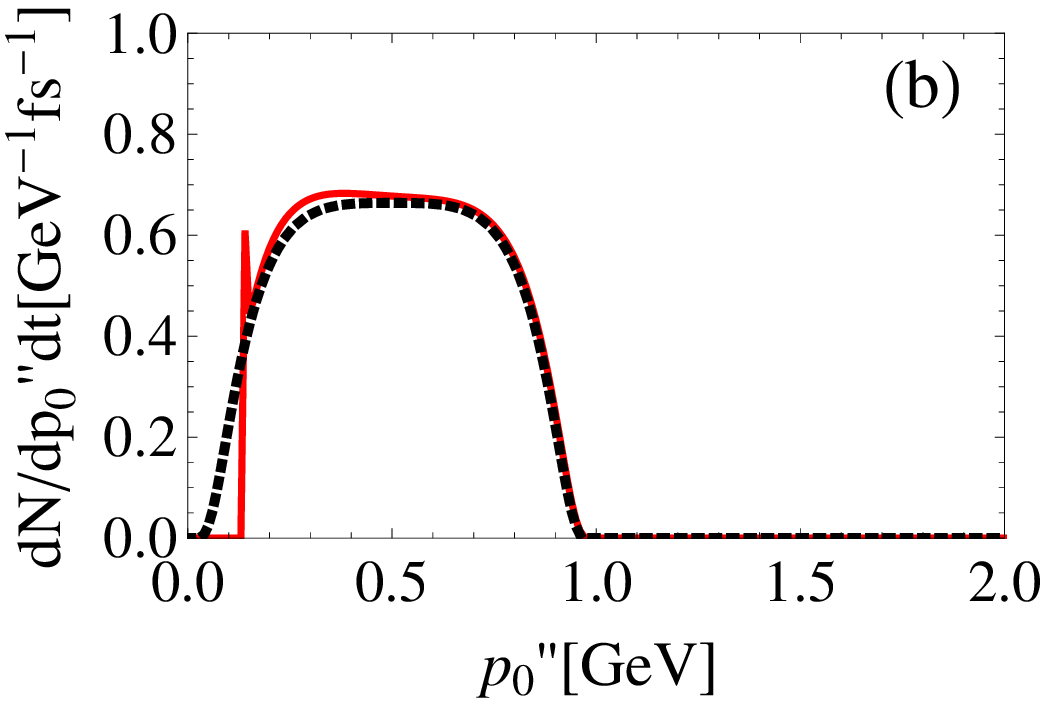} \\
            \includegraphics[width=0.23\textwidth]{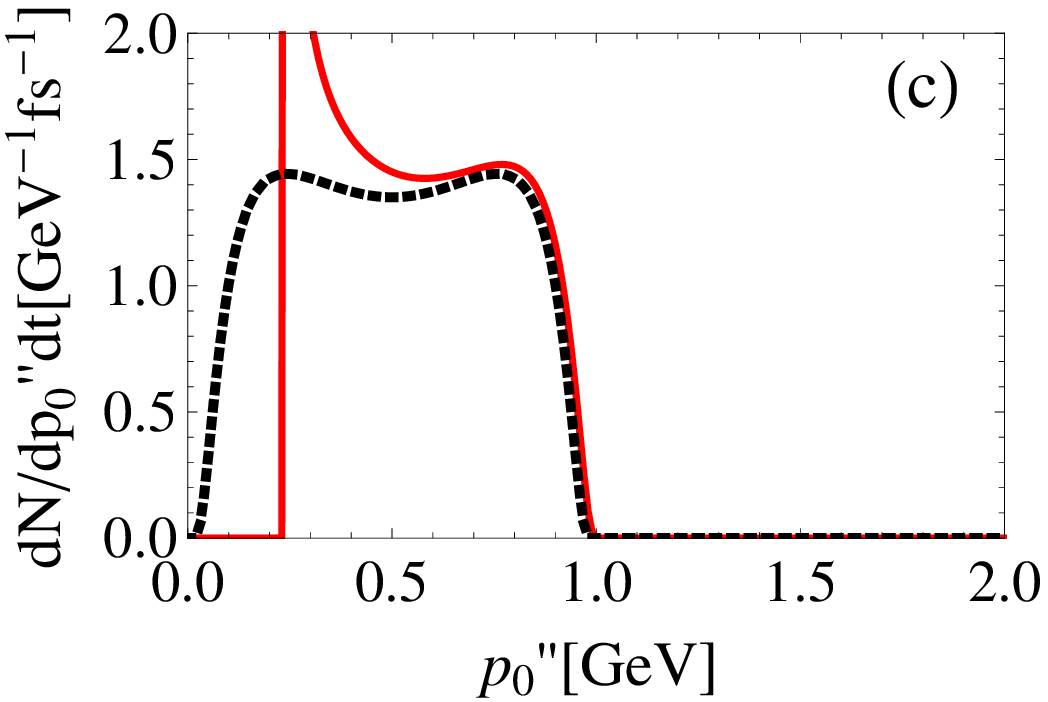}
            \includegraphics[width=0.23\textwidth]{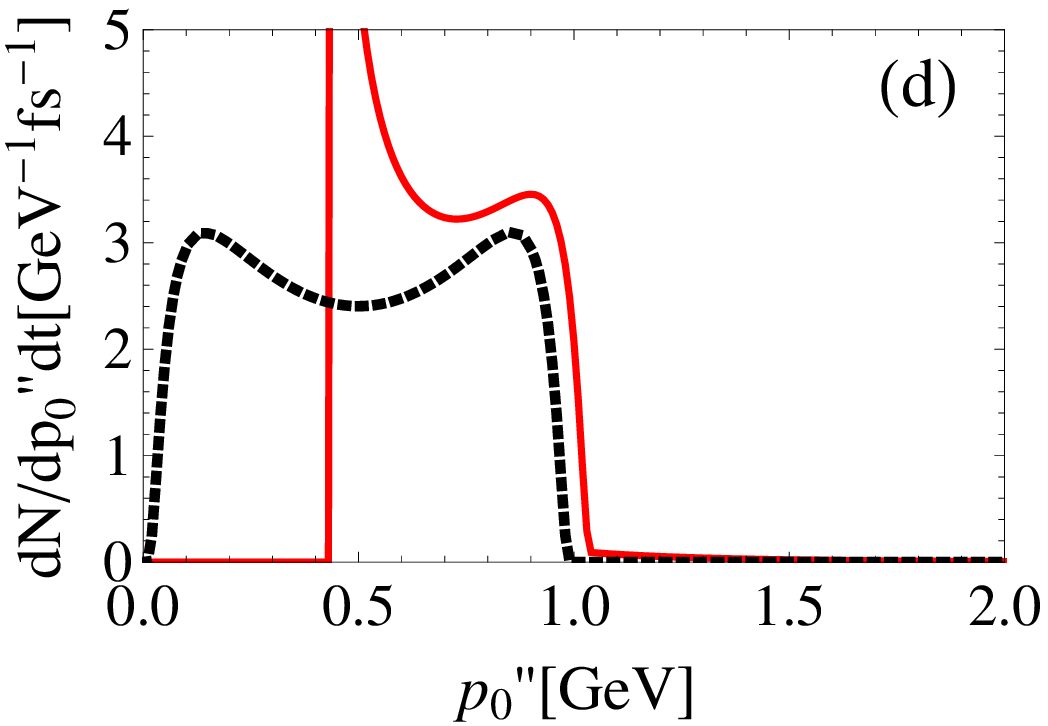}
        \end{center}
        \caption{Multi-photon effects in the emitted electron spectrum $dN_{NBW}/dp''_0 dt$ of a single NBW process in ultra intense lasers (red lines). Parameters are: $\theta=\pi$, $\phi=0$, $k'_0=1$GeV, $k_0=1.24$eV($\lambda=1\mu$m) and $I=a_0^2 1.37\times 10^{18}$W/cm$^2$ is $10^{22}$ (a), $10^{23}$ (b), $3\times 10^{23}$ (c) and $10^{24}$W/cm$^2$ (d), respectively. For comparison, spectra without multi-photon effects were also shown (black dotted).}
        \label{Spectrum_NBW}
    \end{figure}

    Multi-photon effect on single NBW spectrum in an ultra intense laser is a bit different. When laser intensity is comparatively low, the $e^-$ spectrum with multi-photon effect agrees with that without it as Fig. \ref{Spectrum_NBW} (a) shows, note that the spectrum of $e^+$ is the same. Then with the growth of laser intensity, as shown in Fig. \ref{Spectrum_NBW} (b) and (c), the multi-photon effect introduces a lower limit into the spectrum. Finally, when laser intensity is very high, the cutoff at $k'_0$ disappears, which is shown in Fig. \ref{Spectrum_NBW} (d).

\begin{figure}
        \begin{center}
            \includegraphics[width=0.23\textwidth]{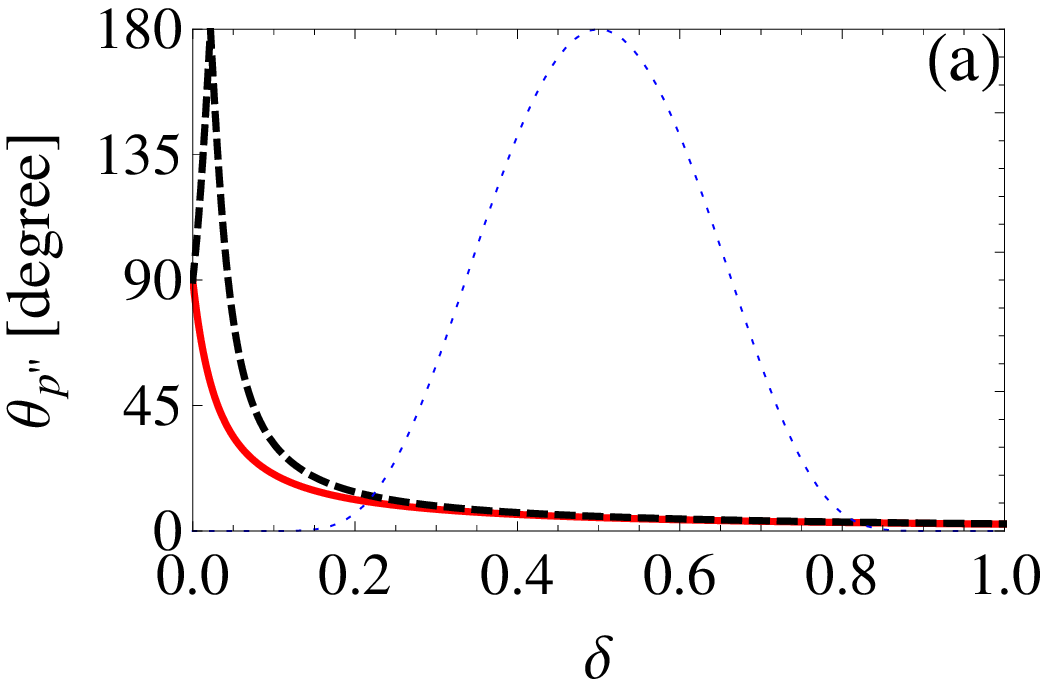}
            \includegraphics[width=0.23\textwidth]{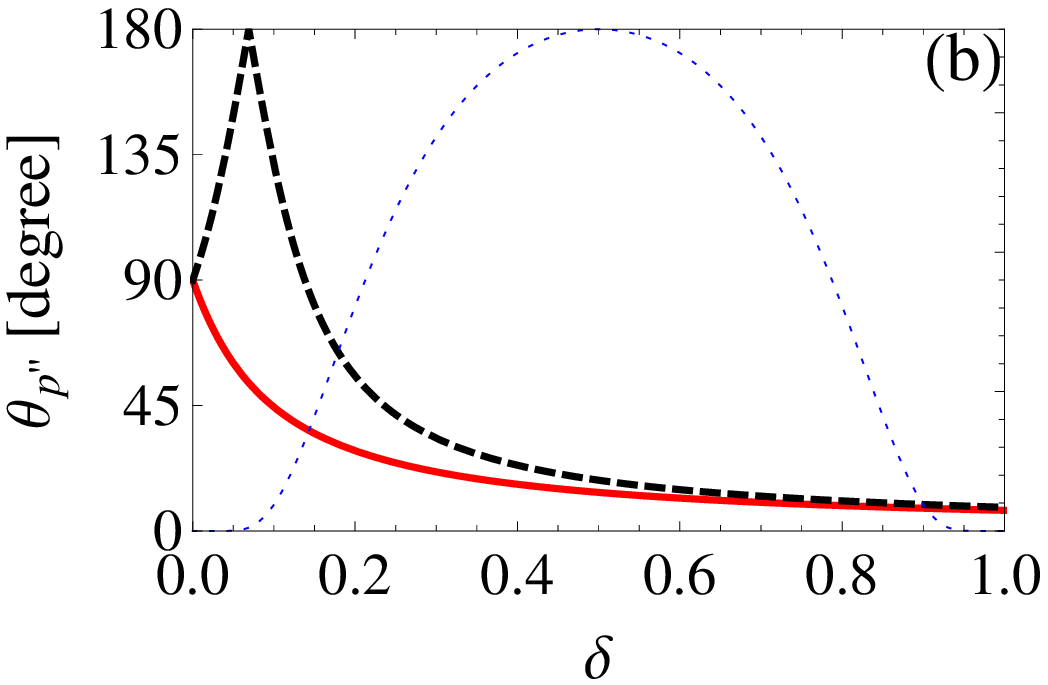} \\
            \includegraphics[width=0.23\textwidth]{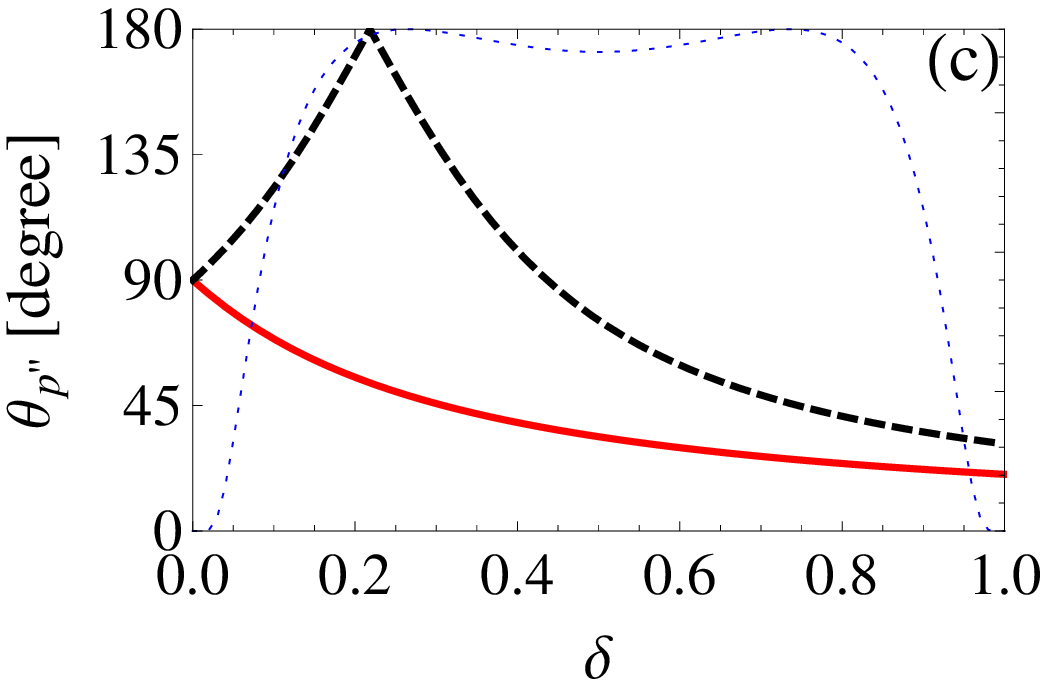}
            \includegraphics[width=0.23\textwidth]{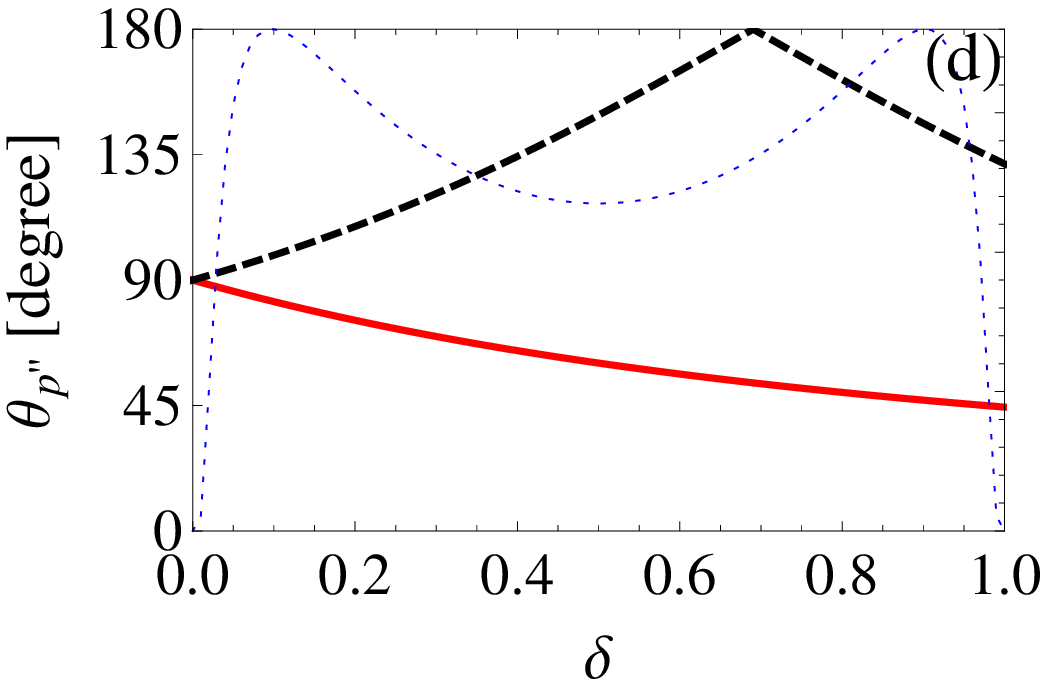}
        \end{center}
        \caption{Emission angle $\theta_{p''\pm}$ (red solid line for $+$ and black dashed for $-$) of NBW with respect to instantaneous $\gamma$ photon momentum $\textbf{k}'$ in ultra intense lasers. Parameters are: $\theta=\pi/2$, $\phi=0$, $k'_0=1$GeV, $k_0=1.24$eV($\lambda=1\mu$m) and $I=a_0^2 1.37\times 10^{18}$W/cm$^2$ is $10^{22}$ (a), $10^{23}$ (b), $10^{24}$ (c) and $10^{25}$W/cm$^2$ (d), respectively. The blue dotted lines are corresponding normalized differential probability.}
        \label{EmissionAngle_NBW}
    \end{figure}

    Different from the fixed emission angle of a single NCS in ultra intense lasers, the non vanishing emission angles of NBW for both $e^-$ and $e^+$ are not fixed. Fig. \ref{EmissionAngle_NBW} shows the emission angle $\theta_{p''\pm}$ of $e^-$ of a single NBW in ultra intense laser strongly depends on $\delta=kp''/kk'$. Note that emission angle $\theta_{p'''\pm}$ of $e^+$ is symmetric with respect to $\delta=0.5$ for the symmetry between $e^-$ and $e^+$ in NBW. The differential probability is also present to show the range of physically significant $\delta$ at different intensity. When laser intensity is low, the probability that $e^-$ and $e^+$ are emitted along directions close to $\textbf{k}'$ is almost $1$. The probability of large angle $e^-$ and $e^+$ emission grows with laser intensity, and becomes dominant when $I \gtrsim 10^{24}$W/cm$^2$.

\section{Signals of Multi-Photon Effects in Nonlinear Compton Scattering and Nonlinear Breit-Wheeler Process in Future Experiments}\label{sec:succs}
    In the last two sections, we have deduced multi-photon effect in single NCS/NBW processes in ultra intense lasers. However, in corresponding experiments that collide an ultra intense laser pulse with an electron beam, back reactions and secondary interactions are very important. NCS processes can happen successively for several times to an electron when it propagates through the pulse, generated $\gamma$ photons would also produce $e^- - e^+$ pairs through NBW. Hence results of single NCS/NBW in ultra intense lasers given in section \ref{sec:single} and \ref{sec:single_NBW} are not easy to directly observe.

    To explore signals of multi-photon effects in NCS and NBW on possible future laser facilities, a Monte-Carlo simulation of interactions between an electron bunch and a laser pulse is carried out. Eq. (\ref{uSpectrum_NPSIncl}), (\ref{GeneralkP1}) and (\ref{GeneralpP1}) are employed to describe NCS and NBW including multi-photon effects is described by Eq. (\ref{NBW_Ritus}), (\ref{GeneralpPP_NBW}) and (\ref{GeneralpPPP_NBW}). Between emissions, classical equations of motion are applied to describe electron propagation. Other processes such as higher order radiations \cite{BenKing2013,BenKing2015}, Schwinger pair production \cite{Schwinger1951}, vacuum birefringence \cite{VacuumBirefringence}, photon splitting \cite{PhotonSplitting} and vacuum radiation \cite{VacRad} are neglected for their much smaller effects under laser conditions concerned.

    In the simulation, a quasi mono-energetic bunch of $1.022$GeV electrons collides head-on with a tightly focused linearly polarized short laser pulse. The laser field is given by an approximate solution of Maxwell's equations to the first order of $(k_0 w_0)^{-1}$ and $(\omega_0 \tau_0)^{-1}$, where $k_0$, $w_0$, $\tau_0$ are the wave vector, waist radius and pulse duration, respectively \cite{FocusedField,LiJianXin2014}. Electromagnetic force between electrons is ignored for it is typically $7-8$ magnitudes weaker than the laser Lorentz force.

    Applied parameters are the following: peak intensity of the laser pulse is $I_0= E_{max}^2/2 = 10^{23}$W/cm$^2$ and $10^{24}$W/cm$^2$, the laser wave length $\lambda=1\mu$m, beam waist $w_0=1\mu$m and duration $\tau_0=2\lambda/c=6.7$fs. The electron bunch includes $10^6$ electrons, which uniformly distribute in a $R=1\mu$m sphere. The mean initial electron energy is $1.022$GeV ($\gamma_0=2000$), and both the energy dispersion $\Delta \gamma/\gamma_0$ and angular dispersion $\Delta \theta$ are $0.001$. The initial distance between the electron bunch and laser pulse center is $3\mu$m, (the simulation lasts $30$fs, which allows most particles to escape the pulse).

    \begin{figure}
        \begin{center}
            \includegraphics[height=0.17\textwidth]{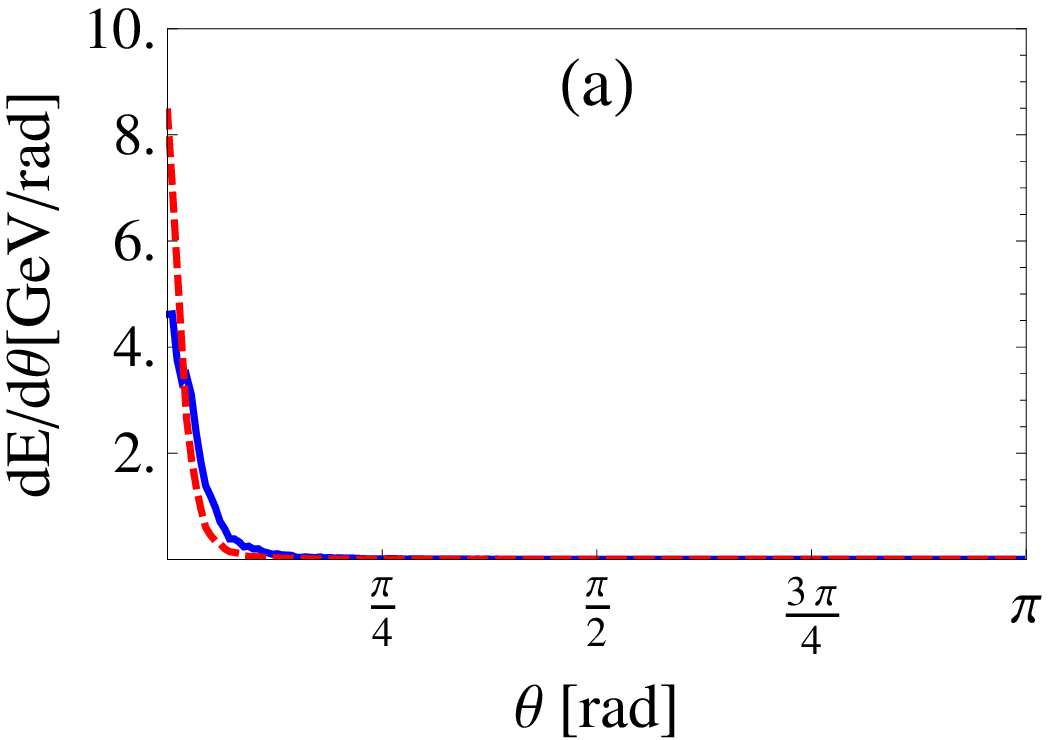}
            \includegraphics[height=0.17\textwidth]{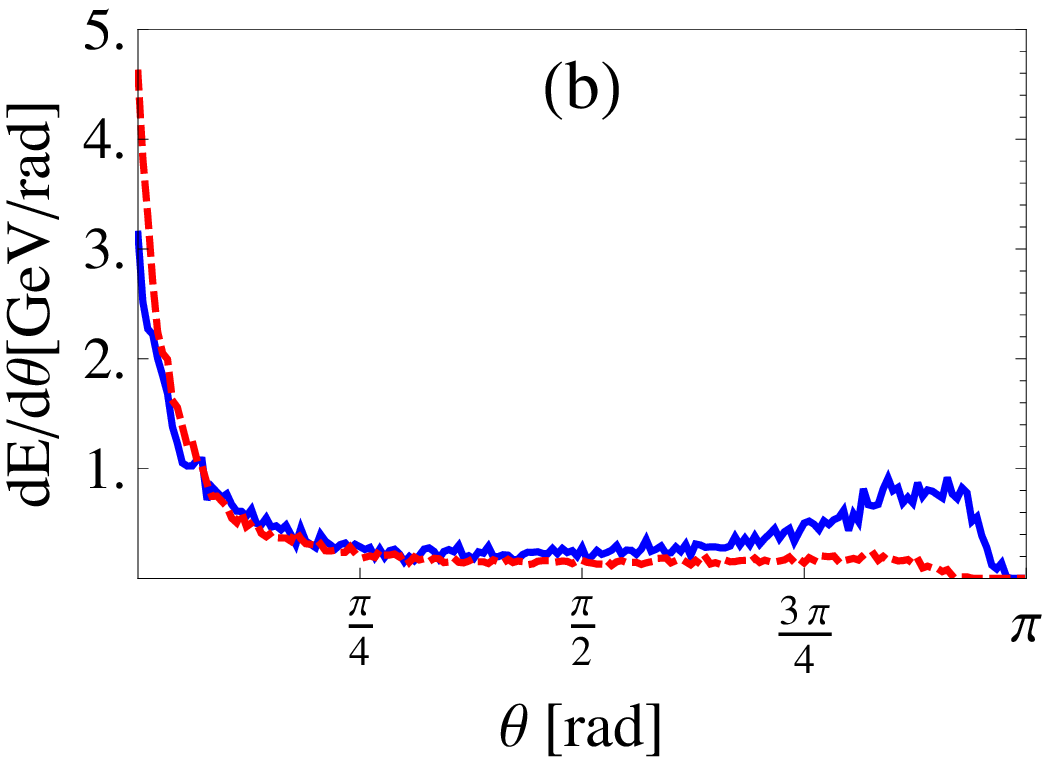} \\
            \includegraphics[height=0.16\textwidth]{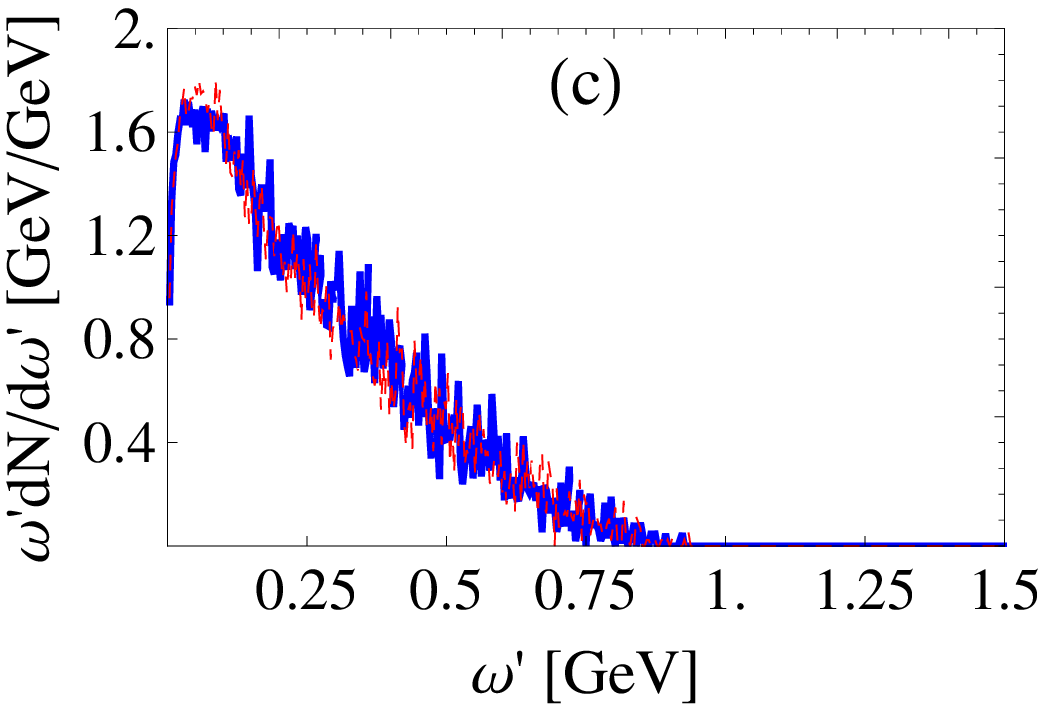}
            \includegraphics[height=0.16\textwidth]{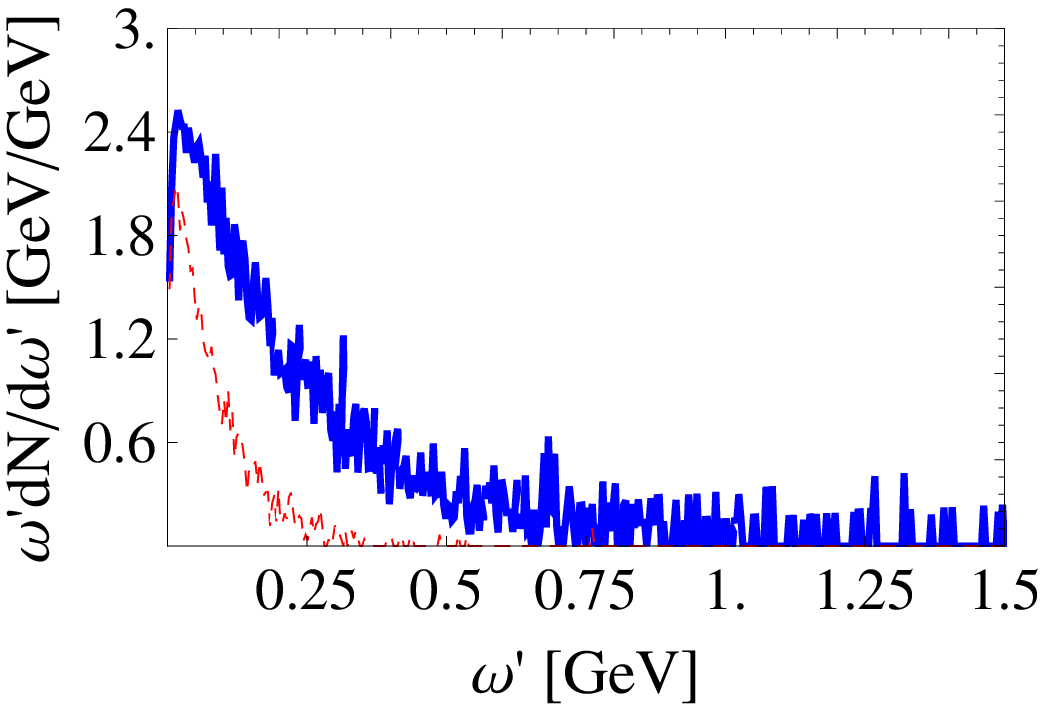}
        \end{center}
        \caption{Angular distribution of radiation intensity for $I_0=10^{23}$W/cm$^2$ (a) and $I_0=10^{24}$W/cm$^2$ (b) and energy integrated spectra of emitted photons for $I_0=10^{23}$W/cm$^2$ (c) and $I_0=10^{24}$W/cm$^2$ (d), (d) only includes photons in the backward hemisphere of initial electron bunch. All normalized by the initial electron number $N_e$, blue solid lines include and red dashed lines exclude multi-photon effects.}
        \label{Simulation_Results}
    \end{figure}

    Simulation result for $I_0=E_{max}^2/2=1 \times 10^{23}$W/cm$^2$ is shown in Fig. \ref{Simulation_Results} (a). Radiation intensity angular distribution of emitted $\gamma$ photons including and excluding multi-photon effect both concentrate around the initial electron bunch direction, besides consecutive stochastic emissions which can broaden the angular distribution \cite{LiJianXin2014,LiJianXin2015}, multi-photon effect further doubles the angular divergence (FWHM) from $3^\circ$ to $6^\circ$.

    When laser peak intensity is increased to $10^{24}$W/cm$^2$, as shown in Fig. \ref{Simulation_Results} (b), multi-photon effect creates a strong, new peak in emission intensity angular distribution which is very close to the opposite direction. Multi-photon effects in the spectrum of backward hemisphere is also significant. Fig. \ref{Simulation_Results} (d) shows that, the spectrum in backward hemisphere excluding multi-photon effects is bounded below $0.5$ GeV while that including multi-photon effects extends $1.5$GeV. The strength of spectrum is also several times stronger in the backward hemisphere.

\section{Conclusion}\label{sec:concl}
    In conclusion, multi-photon effects in NCS/NBW in ultra intense lasers is investigated. Present understanding of single NCS in ultra intense lasers is based on three classical/semi-quantum ideas of ultra relativistic electron radiation, forward emission along instantaneous electron forward direction, recoil reaction along instantaneous electron backward direction and spectrum cutoff at instantaneous electron energy. Single NBW process in ultra intense lasers is also believed to emit along instantaneous $\gamma$ photon forward direction and has spectrum cutoff at instantaneous $\gamma$ photon energy. We show that these ideas are approximations excluding multi-photon effects, which is correct only in comparatively weak fields.

    In ultra intense lasers with $a_0 \gg 1$, when the total momentum of laser photons scattered in a single NCS/NBW process in ultra intense laser becomes comparable to that of the electron/$\gamma$ photon itself, NCS/NBW can have large emission angles and even backward. Phenomena accompany large angle emission of NCS are strong deflection of electron instead of a recoil right backward and the spectrum surpasses instantaneous electron energy. For NBW, the spectrum gets an additional lower limit.

    Simulation results of successive NCS and NBW demonstrate that multi-photon effects would dominate possible future SFQED experiments that collide GeV scale electron bunches with $10^{24}$W/cm$^2$ ultra intense laser pulses. Multi-photon effects greatly enhance backward emission, including create a backward emission peak and very high energy photons above initial electron energy. Additionally, although multi-photon effect is weaker at $10^{23}$W/cm$^2$, its signal is still strong enough for measurement.

\acknowledgments
    The author would like to thank Dr. Benjamin King and Dr. Huayu Hu for helpful discussions. This work is supported by the Science Challenge Project (Grant No. TZ2016005) and the Foundation of Science and Technology on Plasma Physics Laboratory, China (Grant No. 9140C680303140C68288).

\end{document}